\documentclass[%
 reprint,
 superscriptaddress,
nobibnotes,
amsmath,amssymb,
aps,
prl,
longbibliography,
showkeys,
]{revtex4-2}

\usepackage{graphicx}
\usepackage{dcolumn}
\usepackage{bm}
\usepackage[export]{adjustbox}
\usepackage{xcolor}
\usepackage{booktabs}
\usepackage{multirow}
\usepackage{siunitx}

\usepackage{hyperref}

\begin{document}

\setcitestyle{super}

\title{Observation of Aerosolization-induced Morphological Changes in Viral Capsids}

\author{Abhishek~Mall}
\affiliation{Max Planck Institute for the Structure and Dynamics of Matter, 22761 Hamburg, Germany}

\author{Anna~Munke}
\affiliation{Center for Free Electron Laser Science, Deutsches Elektronen Synchrotron (DESY), 22607 Hamburg, Germany}
\affiliation{Laboratory of Molecular Biophysics, Department of Cell and Molecular Biology, Uppsala University, Uppsala, SE-75124, Sweden}

\author{Zhou~Shen}
\affiliation{Max Planck Institute for the Structure and Dynamics of Matter, 22761 Hamburg, Germany}

\author{Parichita~Mazumder}
\affiliation{Max Planck Institute for the Structure and Dynamics of Matter, 22761 Hamburg, Germany}

\author{Johan~Bielecki}
\affiliation{European XFEL, Holzkoppel 4, 22869 Schenefeld, Germany}

\author{Juncheng~E}
\affiliation{European XFEL, Holzkoppel 4, 22869 Schenefeld, Germany}

\author{Armando~Estillore}
\affiliation{Center for Free Electron Laser Science, Deutsches Elektronen Synchrotron (DESY), 22607 Hamburg, Germany}

\author{Chan~Kim}
\affiliation{European XFEL, Holzkoppel 4, 22869 Schenefeld, Germany}

\author{Romain~Letrun}
\affiliation{European XFEL, Holzkoppel 4, 22869 Schenefeld, Germany}

\author{Jannik~L{\"u}bke}
\affiliation{Center for Free Electron Laser Science, Deutsches Elektronen Synchrotron (DESY), 22607 Hamburg, Germany}

\author{Safi~Rafie-Zinedine}
\affiliation{European XFEL, Holzkoppel 4, 22869 Schenefeld, Germany}
\affiliation{Institute of Biomaterials and Biomolecular Systems, University of Stuttgart, Pfaffenwaldring 57, 70569 Stuttgart, Germany}

\author{Adam~Round}
\affiliation{European XFEL, Holzkoppel 4, 22869 Schenefeld, Germany}

\author{Ekaterina~Round}
\affiliation{European XFEL, Holzkoppel 4, 22869 Schenefeld, Germany}

\author{Michael~R{\"u}tten}
\affiliation{Universit{\"a}t Hamburg, Institute of Physical Chemistry, Grindelallee 117 20146 Hamburg, Germany}

\author{Amit~K.~Samanta}
\affiliation{Center for Free Electron Laser Science, Deutsches Elektronen Synchrotron (DESY), 22607 Hamburg, Germany}
\affiliation{The Hamburg Center for Ultrafast Imaging, Universität Hamburg, 22761 Hamburg, Germany}

\author{Abhisakh~Sarma}
\affiliation{European XFEL, Holzkoppel 4, 22869 Schenefeld, Germany}

\author{Tokushi~Sato}
\affiliation{European XFEL, Holzkoppel 4, 22869 Schenefeld, Germany}

\author{Florian~Schulz}
\affiliation{Universit{\"a}t Hamburg, Institute of Physical Chemistry, Grindelallee 117 20146 Hamburg, Germany}

\author{Carolin~Seuring}
\affiliation{Centre for Structural Systems Biology (CSSB), Notkestraße 85, 22607 Hamburg, Germany}

\author{Tamme~Wollweber}
\affiliation{Max Planck Institute for the Structure and Dynamics of Matter, 22761 Hamburg, Germany}

\author{Lena~Worbs}
\affiliation{Center for Free Electron Laser Science, Deutsches Elektronen Synchrotron (DESY), 22607 Hamburg, Germany}
\affiliation{Department of Physics, Universität Hamburg, Luruper Chaussee 149, 22761 Hamburg, Germany}

\author{Patrik~Vagovic}
\affiliation{European XFEL, Holzkoppel 4, 22869 Schenefeld, Germany}

\author{Richard~Bean}
\affiliation{European XFEL, Holzkoppel 4, 22869 Schenefeld, Germany}

\author{Adrian~P.~Mancuso}
\thanks{Present address: Diamond Light Source, Harwell Science and Innovation Campus, Didcot, Oxfordshire, OX11 0DE, UK}
\affiliation{European XFEL, Holzkoppel 4, 22869 Schenefeld, Germany}
\affiliation{Department of Chemistry and Physics, La Trobe Institute for Molecular Science, La Trobe University, Melbourne, VIC, 3086, Australia}

\author{Ne-Te~Duane~Loh}
\affiliation{National University of Singapore (NUS), Dep. of Physics / Fac. of Science , 2 Science Drive 3 Singapore 117542 , Singapore}

\author{Tobias~Beck}
\affiliation{Universit{\"a}t Hamburg, Institute of Physical Chemistry, Grindelallee 117 20146 Hamburg, Germany}

\author{Jochen~K{\"u}pper}
\affiliation{Center for Free Electron Laser Science, Deutsches Elektronen Synchrotron (DESY), 22607 Hamburg, Germany}
\affiliation{The Hamburg Center for Ultrafast Imaging, Universität Hamburg, 22761 Hamburg, Germany}
\affiliation{Department of Physics, Universität Hamburg, Luruper Chaussee 149, 22761 Hamburg, Germany}
\affiliation{Department of Chemistry, Universität Hamburg, 20146 Hamburg, Germany}

\author{Filipe~R.N.C.~Maia}
\affiliation{Laboratory of Molecular Biophysics, Department of Cell and Molecular Biology, Uppsala University, Uppsala, SE-75124, Sweden}
\affiliation{NERSC, Lawrence Berkeley National Laboratory, Berkeley, CA, 94720, United States}

\author{Henry~N.~Chapman}
\affiliation{Center for Free Electron Laser Science, Deutsches Elektronen Synchrotron (DESY), 22607 Hamburg, Germany}
\affiliation{Department of Physics, Universität Hamburg, Luruper Chaussee 149, 22761 Hamburg, Germany}
\affiliation{The Hamburg Center for Ultrafast Imaging, Universität Hamburg, 22761 Hamburg, Germany}

\author{Kartik~Ayyer}
\email{kartik.ayyer@mpsd.mpg.de}
\affiliation{Max Planck Institute for the Structure and Dynamics of Matter, 22761 Hamburg, Germany}
\affiliation{The Hamburg Center for Ultrafast Imaging, Universität Hamburg, 22761 Hamburg, Germany}



\maketitle

\textbf{Single-stranded RNA viruses co-assemble their capsid with the genome and variations in capsid structures can have significant functional relevance~\cite{sun2010genome,jana2020structural}. In particular, viruses need to respond to a dehydrating environment to prevent genomic degradation and remain active upon rehydration. Theoretical work has predicted low-energy buckling transitions in icosahedral capsids which could protect the virus from further dehydration~\cite{widom2007soft}. However, there has been no direct experimental evidence, nor molecular mechanism, for such behaviour. Here we observe this transition using X-ray single particle imaging of MS2 bacteriophages after aerosolization~\cite{seibert2011single,bielecki2019electrospray}. Using a combination of machine learning tools, we classify hundreds of thousands of single particle diffraction patterns to learn the structural landscape of the capsid morphology as a function of time spent in the aerosol phase. We found a previously unreported compact conformation as well as intermediate structures which suggest an incoherent buckling transition which does not preserve icosahedral symmetry. Finally, we propose a mechanism of this buckling, where a single 19-residue loop is destabilised, leading to the large observed morphology change~\cite{brodmerkel2022stability}. Our results provide experimental evidence for a mechanism by which viral capsids protect themselves from dehydration. In the process, these findings also demonstrate the power of single particle X-ray imaging and machine learning methods in studying biomolecular structural dynamics.}

Viral capsids assemble optimally to prioritise the protection and efficient packaging of the genome. It ensures the survival of the virus and facilitates interactions with a host to maintain infectivity. Most spherical viruses in nature assemble their capsids with icosahedral symmetry, characterised by a triangulation number (T): the number of structural subunits forming the triangular facets of the icosahedron~\cite{caspar1962physical}. For instance, the MS2 bacteriophage, a 27 nm single-stranded RNA virus infecting \emph{Escherichia coli} bacteria (\emph{E. coli}), is a non-enveloped virus with a T = 3 icosahedral capsid structure~\cite{valegaard1990three}. With non-genomic RNA, the capsid protein can also assemble into T = 4 as well as hybrid capsids between these two triangulation numbers~\cite{de2020bacteriophage}. Furthermore, covalent dimerization of the coat protein in MS2 can lead to an octahedral structure under certain buffer conditions~\cite{plevka2008crystal}.

The variability in capsid structures and symmetry breaking in icosahedral capsids can potentially affect infectivity, and has been well-studied in the context of viral maturation~\cite{jana2020structural}. The shape of the capsids is determined by elastic properties such as stretching and bending energies, spontaneous curvature, and chirality. The transition from smooth to faceted shapes in icosahedral capsid shells corresponds to a soft-mode buckling transition, driven by bending stiffness~\cite{widom2007soft}. Continuum elasticity theory attributes shape transitions in capsids with non-icosahedral symmetries to a trade-off between stretching and bending energies~\cite{nguyen2005elasticity}. Moreover, the elastic responses to external forces elucidate the mechanical stability and rupture behaviour of both empty and filled viral capsids~\cite{buenemann2008elastic}.

\begin{figure*}
\centering
\includegraphics[scale=1]{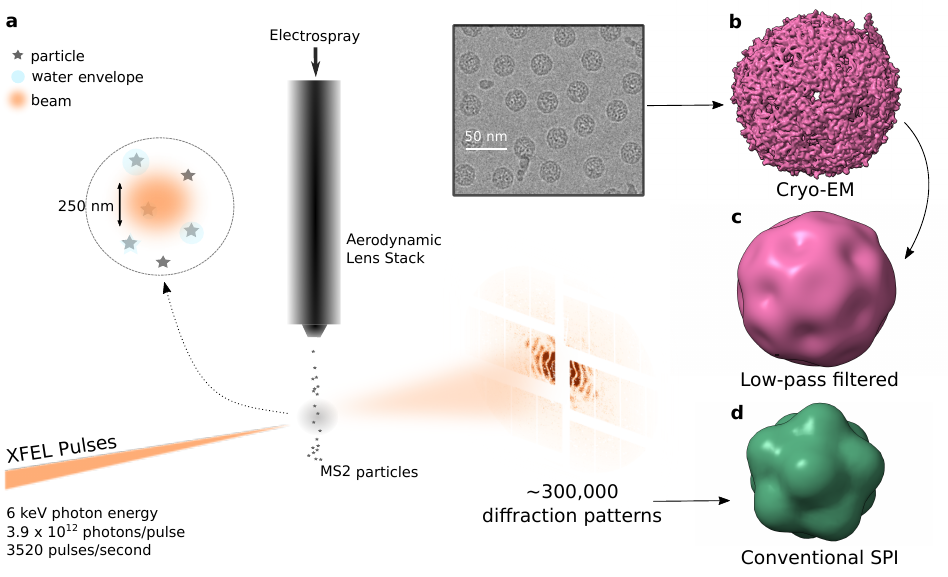} \\
\caption{\textbf{X-ray SPI Experiment}. (a) MS2 bacteriophage particles, $\sim$ 27 nm in diameter, were aerosolized using an electrospray and focused with an aerodynamic lens stack to the interaction region within the X-ray beam of $250\times$\SI{250}{\nano\meter\squared} focus. The top inset shows a representative cryo-electron microscopy (cryo-EM) micrograph of the particles. (b) The 3D structure of MS2 capsid determined by cryo-EM (resolution \SI{0.49}{\nano\meter}), served as the control for the subsequent X-ray SPI experiment. (c) The same cryo-EM structure was low-pass filtered to the resolution of the conventional X-ray SPI structure. (d) The structure retrieved from diffraction data (6.1 nm resolution) using the conventional analysis pipeline is notably different from the cryo-EM structure in (c).}
\label{fig:SPI} 
\end{figure*} 

Understanding the intricate and non-trivial variations in viral capsid structure is essential for unravelling the fundamental processes driving viral infectivity and hardiness. In this study, we approach this problem using the emerging technique of single particle imaging (SPI) at an X-ray Free Electron Laser (XFEL) source. This is a powerful method for probing the structures of nanoscale systems~\cite{seibert2011single,ayyer20213d,shen2024resolving}. In these experiments, extremely bright, ultrashort and coherent X-ray pulses from XFELs interact with copies of isolated single particles in random orientations one at a time. This process generates millions of diffraction patterns, each from a single viral particle. Machine learning approaches, including unsupervised methods~\cite{zhuang2022unsupervised,ignatenko2021classification}, are employed to identify diffraction patterns scattered from the target object amid contaminants, aggregates and outliers. This is followed by orientation determination and phase retrieval to obtain the electron density of the average particle~\cite{ayyer2016dragonfly,ayyer2019low,rose2018single}. Since each measurement is made on an individual particle, one additionally has the opportunity to classify them and obtain not only the average structure, but also the landscape of structural variations~\cite{ayyer20213d,cho2021high,zhuang2022unsupervised,shen2024resolving}.

The short pulses of an XFEL also enable time-resolved SPI experiments to investigate ultrafast phenomena and structural dynamics in ensembles of particles at the nanoscale. This progression has enabled the exploration of ultrafast photo-induced dynamics~\cite{hoeing2023time}, resolving the non-equilibrium shape distributions~\cite{shen2024resolving}, retrieving the 3D morphology of polyhedral particles~\cite{colombo2023three}, melting to explosive disintegration of nanoparticles~\cite{dold2023melting}, demonstrating diffraction before destruction at the protein scale~\cite{ekeberg2024observation} and retrieving structures of heterogeneous nanoparticles~\cite{ayyer20213d}. 

In this work, we explore and analyse the structural dynamics of MS2 bacteriophage viruses after aerosolisation. In the process of being transported to the XFEL beam, the aerosol droplets are continuously drying, simulating the natural dehydration process~\cite{thomas2004electrospray,martin2018exploring,coleman2024effect}. The particles are then probed using the XFEL at random degrees of dehydration to produce single particle diffraction patterns. Using a combination of maximum likelihood and deep learning techniques, we map the collected diffraction data from the ensemble of MS2 capsids to a continuous structural landscape. One can then observe viral capsid structures ranging from the fully-hydrated state to a previously unobserved capsid form with full coverage of intermediate structures. This data then enabled us to hypothesise a molecular mechanism for the observed conformational changes, which seems to apparently protect the genome from further dehydration. In the process, we also show how the combination of machine learning methods with high-throughput SPI measurements at XFELs can be used to understand the conformational landscape and dynamics of biomolecules in a fairly general manner. 

\section{X-ray SPI Experiment}
MS2 bacteriophage particles in an aqueous buffer were aerosolised and sequentially injected into the X-ray beam interaction region using an electrospray-ionisation aerodynamic-lens-stack sample delivery system~\cite{bielecki2019electrospray}, as shown in Fig.~\ref{fig:SPI}a. Diffraction patterns were collected at an average rate of 3520 frames/second for an integrated collection time of \SI{3.6}{\hour} with a hit ratio of around $0.7 \%$. Frames with diffraction from particles were detected by setting a threshold on the scattered signal. A total of \num{287168} potential hit diffraction patterns were identified containing 1350 photons per pattern on average.

The highly noise-tolerant EMC algorithm~\cite{loh2009reconstruction} can be used to categorise and orient diffraction frames with only a few photons~\cite{ayyer2019low,philipp2012solving,giewekemeyer2019experimental}. We employed the \emph{Dragonfly} software~\cite{ayyer2016dragonfly}, to perform two-dimensional (2D) classification using this algorithm. This procedure generated multiple 2D intensity models of diffraction patterns in the detector plane~\cite{ayyer20213d} by determining the in-plane rotation angle and relative incident fluence of each diffraction pattern. These 2D reciprocal space intensity models capture the average of aligned copies of a subset of patterns from the entire dataset, following which class averages were manually selected corresponding to single particles, indicated by high fringe contrast and a convex envelope. This procedure was then repeated, each time rejecting the various contaminants like aggregates and other outliers. 

The final selection contained intensity models revealing distinctive diffraction features corresponding to an icosahedral particle with good contrast and sharp streaks. The subset of diffraction frames associated with this intensity model was selected to reconstruct a three-dimensional (3D) Fourier model using \emph{Dragonfly}. Icosahedral symmetrization was applied due to the subset having only \num{7249} patterns. Subsequently, it was phased to retrieve the  electron density of the MS2 capsid, as shown in Fig.~\ref{fig:SPI}d, with an estimated resolution of \SI{6.1}{\nano\meter}. 

Even at this resolution, this structure is markedly different from the one obtained using cryo-electron microscopy (cryo-EM) on the same sample batch shown in Fig.~\ref{fig:SPI}b. This structure at \SI{0.49}{\nano\meter} resolution provides insight into the conformation of the hydrated, flash-frozen capsid,  which is a near-spherical particle with icosahedral symmetry (see Methods section Cryo-EM Structure Determination for details). This baseline structure, along with the similar crystallographic structure of the capsid (PDB: 2MS2)~\cite{golmohammadi1993refined} serves as a reference for interpreting structural changes induced by aerosolisation. The low-pass filtered version shown in Fig.~\ref{fig:SPI}c shows differences at both the 5-fold and 3-fold sites, with the X-ray structure indenting inwards at the 3-fold sites.

The fact that only a limited number of patterns (only 2.5\%) went into the final 3D structure with the conventional X-ray SPI analysis pipeline raises the question of the structures of the rejected particles and the source of the heterogeneity. As the particles traverse through the low-humidity environment aerodynamic lens and then the vacuum environment of the interaction region, the surrounding water envelope is continuously evaporating. We explore the possibility of whether the rejected patterns contain information about the transition from the hydrated state to the final structure depicted in Fig.~\ref{fig:SPI}d.

\section{Heterogeneity Analysis Workflow}

\begin{figure}
\includegraphics[width=0.95\columnwidth]{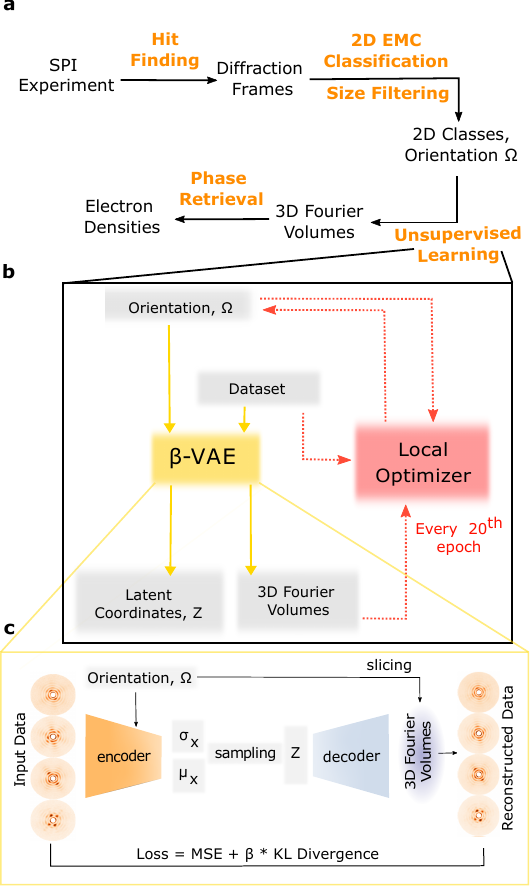} \\
\caption{\textbf{Analysis Pipeline.} (a) Schematic of diffraction data analysis workflow. All steps other than the unsupervised learning ones are part of the standard SPI workflow. (b) Detailed flow chart of the unsupervised learning step to generate the structural latent space and associated 3D Fourier intensity volumes. The pipeline involves training a $\beta$-VAE with a local orientation optimizer. Every 20$^{th}$ epoch, the optimizer outputs an improved estimate for the orientation ($\Omega$) given the 3D Fourier volumes, dataset, and the current estimate of the orientations. (c) Schematic representation of the $\beta$-VAE network. The model takes 2D class-average intensities and orientations as input and encodes them into a latent space via an encoder network. This latent space coordinate (Z) is subsequently utilized by a decoder network to reconstruct 3D Fourier volumes.}
\label{fig:VAE} 
\end{figure}

Figure~\ref{fig:VAE}a shows the analysis workflow for learning the structural landscape of aerosolized MS2 capsids. By using a much larger fraction of the data, we can reconstruct not just a single homogeneous object, but a whole family of structures, and then to study the variations in that family. We first used the same 2D classification approach as for the single reconstruction above. In order to effectively train and utilise the deep learning method discussed below, we expanded the total number of intensity models by performing multiple runs of 2D classification. In each of the 100 independent bootstrapping runs, $20\%$ of the diffraction frames (from a total of 170355) were randomly selected and classified into 100 distinct 2D intensity models, resulting in \num{10000} intensity models.

Upon scrutinising the 2D intensity models, distinctive patterns emerged, including some with strong streaks in the detector plane from faceted particles but also nearly circular diffraction rings from rounded objects. These observations hinted at particle shapes spanning from icosahedral to almost perfectly round. We applied size filtering on the 2D intensity dataset to retrieve distribution of different discrete heterogeneity in the MS2 particles. The effective size of the particles was determined from each intensity average using a spherical diffraction model~\cite{daurer2017experimental} (see Supplementary Section I). 

We curated a dataset of \SI{2558} 2D intensity models from 79711 diffraction frames, representing particles with different capsid morphologies, but excluding models from dimers, aggregates and other contaminants whose nominal size was outside the 23-31~nm size range. To understand the structural landscape of the remaining particles, we employed an unsupervised deep learning approach -- a variational autoencoder (VAE) network~\cite{kingma2013auto}. Inspired by the pioneering cryoDRGN approach using a VAE network to study heterogeneity in single-particle cryo-EM datasets~\cite{zhong2021cryodrgn} and our prior work on continuous shape transitions in gold nanoparticles~\cite{zhuang2022unsupervised}, we adapted the network for our MS2 virion dataset as a $\beta$-VAE, as illustrated in Fig.~\ref{fig:VAE}c. Details of the network architecture are described in the Online Methods.

\begin{figure*}
\centering
\includegraphics[scale=1]{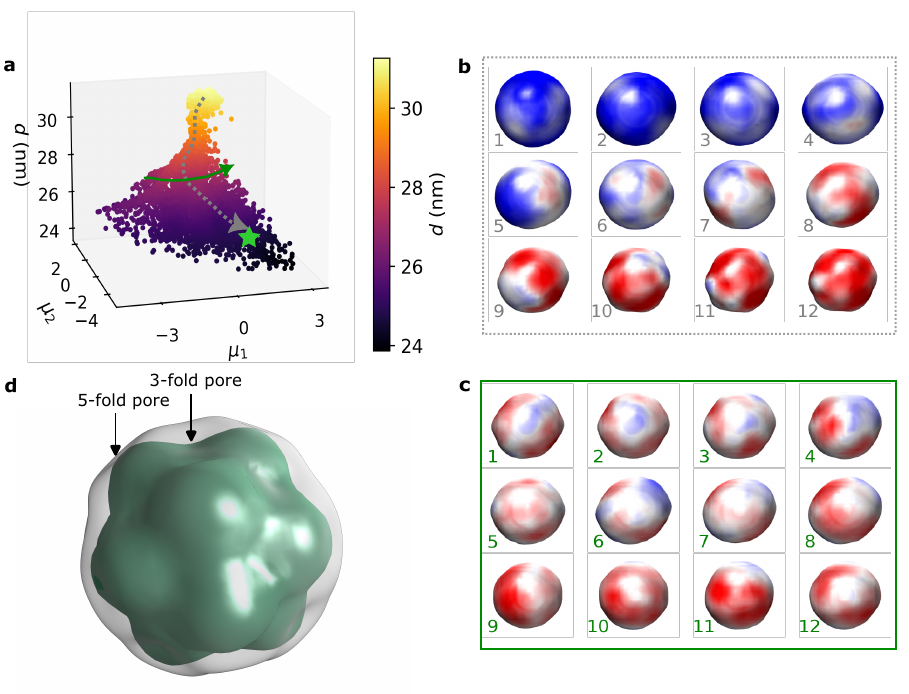}
\caption{\textbf{Structural Landscape}. (a) The latent space learned by the $\beta$-VAE coloured by the estimated diameter (\textit{d}) of individual patterns. The plot highlights two distinct trajectories selected to capture the structural variation phenomenon within the latent space. The retrieved electron density of MS2 particles via phase retrieval of Fourier volumes generated by decoder network of $\beta$-VAE network is shown in the grid. (b) The dotted\textit{Gray} trajectory, following from top to bottom in (a), depicts the shape-size variation in the ensemble of capsids. (c) The \textit{Green} trajectory, progressing from left to right in (a), corresponds to different shape realisations for a fixed size of capsid. Red-white-blue coloured radially. (d) Encapsulated overlay of the 3D structure of the MS2 capsid from the low-pass filtered cryo-EM reconstruction (gray) and the dehydrated X-ray SPI reconstruction (green). The overlap highlights the altered conformations in the vicinity of the 5-fold and 3-fold sites.}
\label{fig:vae_traj} 
\end{figure*}

Conventional autoencoders aim to learn a representation of the dataset by reconstructing the input data. However, the VAE here must learn to separate the effect of true structural variations from the large, but trivial variations in diffraction patterns of differently oriented particles. By incorporating the orientation estimate ($\Omega$) explicitly in the network architecture, we are able to learn the structural variations separately (see Fig.~\ref{fig:VAE}c). The description of the network architecture as well as the analysis workflow to refine the orientation estimates is detailed in Fig.~\ref{fig:VAE}b and the Methods section.

Once trained, the $\beta$-VAE network enables detailed analysis and systematic exploration of structural heterogeneity by examining the 3D intensity volumes reconstructed by the decoder for various points in the latent space. The effective diameter of each 3D volume was determined by fitting a sphere model, the result of which is shown in Fig.~\ref{fig:vae_traj}a where the two components of the latent vector mean, $\mu_1$ and $\mu_2$ are represented along the axes and the colour and height represent the effective diameter. 

\section{Structural landscape}

We highlight two paths through the structural landscape shown in Fig.~\ref{fig:vae_traj}a, capturing two salient features of the evolution of the capsid morphology. Firstly, the dotted \textit{Gray} line trajectory illustrates the variation in shape and size, as observed from top to bottom, which we ascribe to the effect of dehydration. Three-dimensional structures of the MS2 capsids along this path are depicted in Fig.~\ref{fig:vae_traj}b. Following this path, we note a transition from larger particles to smaller, nearly icosahedral particles as dehydration progresses. The largest particles (29-\SI{31}{\nano\meter}) were nearly spherical and larger than the reported 27~nm hydrated structure, representing MS2 capsids with a water envelope around them. The contour images in Fig.~\ref{fig:vae_traj}b and c are colour-coded radially from red to white to blue in order to ease visualisation of facets, curvature, and size changes. The structure in Fig.~\ref{fig:SPI}d, reconstructed without the VAE, lies at the end of this path and is highlighted by a star in the landscape.

The second, \textit{Green}, trajectory is shown in Fig.~\ref{fig:vae_traj}c. Along this path, all particles had an estimated diameter of \SI{27}{\nano\meter}. Here, subtle and gradual deviations from the icosahedral shape at a constant size are observed. Close examination shows structures with varying degrees of deviation from the symmetric structure, also borne out by individual diffraction patterns and class averages showing asymmetric structures. This suggests that whatever morphological change is occurring, is not a coherent change acting on all icosahedral sites simultaneously, but seems to occur independently at each site.

\begin{figure*}
\includegraphics[scale=1]{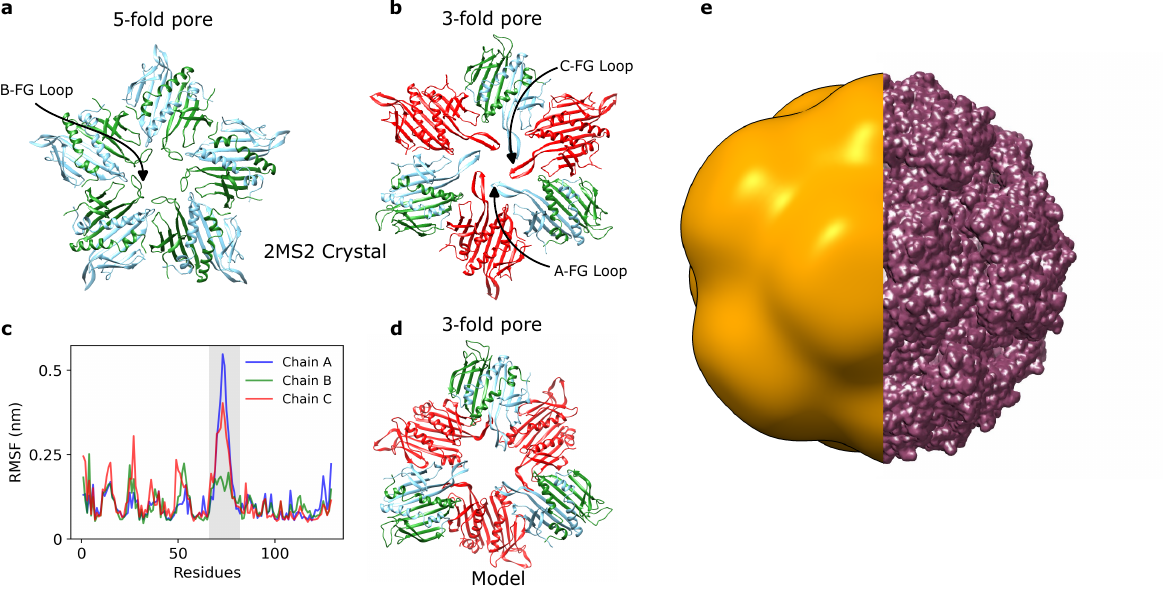} \\
\caption{\textbf{Dehydrated capsid model}.  (a-b) The pentameric (5-A/B) and hexameric (3-A/B and 3-C/C) faces of the T=3 icosahedral capsid shell from the 2MS2 PDB structure. At the 5-fold axis FG loops of B (B-FG loop, green, a), and at the 3-fold axis FG-loops of A (A-FG loop, sky blue, b) and C (C-FG loop, red) are crucial for capsid assembly and curvature. (c ) The A-FG loop and C-FG loop exhibit significant fluctuations compared to B-FG loop (residues 66-82, shaded region). The Root Mean Square fluctuation (RMSF) was calculated from a 20 ns vacuum MD-trajectory of A/B and C/C dimers. (d) Transformed hexameric building block designed/modelled from the X-ray SPI map. At the 3-fold axis, C/C dimers move toward the capsid centre. (e) Map generated from transformed capsid model (at 6.1 nm resolution). The left half is in a similar representation as the experimental X-ray SPI map (Fig.~\ref{fig:SPI}d) for visual comparison.}
\label{fig:Capsid} 
\end{figure*} 

\section{Proposed molecular mechanism}

In order to better understand the capsid morphology change, we focus on the fully dehydrated state and compare it to the MS2 capsid structures obtained from cryo-EM and crystallography. Figure~\ref{fig:vae_traj}d displays an overlay visualisation of the two MS2 capsid structures: the low-pass filtered cryo-EM reconstruction and the X-ray SPI reconstruction. This overlay emphasises the locations of the pores at the 5-fold and 3-fold sites (vertex and face centre respectively), which are affected during dehydration through aerosolisation. The T=3 icosahedral capsid of MS2 consists of 12 5-fold contacts at the vertices and 20 6-fold contacts at the face-centres, as seen in the cryo-EM structures in Fig.~\ref{fig:SPI}b and c. The configuration of the coat protein creates a capsid shell featuring 32 pores (about 2 nm in diameter), denoted here as 5-fold and 3-fold pores, respectively.

The crystal structure of the MS2 virus capsid~\cite{golmohammadi1993refined} shows that the coat protein has three possible conformations, termed A, B and C. These proteins assemble into two types of dimers: asymmetric A/B dimers and symmetric C/C dimers. Although the A, B, and C subunits (129 residues) are almost structurally identical, they differ  in the conformation of the FG-loop (residues 66-82), with the A and C subunits exhibiting a conformation that is different from that of the B subunit.

The 5-fold pores consist of 5 A/B dimers, with the FG-loops of the five B-subunits oriented towards the pores in a compact conformation, as depicted in Fig.~\ref{fig:Capsid}a. The 3-fold pores are formed by six dimers--3 A/B and 3 C/C--arranged alternately, with 3 FG-loops from each A and C subunit in an extended conformation (Fig.~\ref{fig:Capsid}b). The FG-loop plays a pivotal role in capsid assembly and affects its curvature and mutations in this region can disrupt assembly~\cite{ni1995crystal}. 

We performed molecular dynamics (MD) simulations of the A/B and C/C dimers in vacuum conditions similar to those during sample delivery of the SPI experiment. The FG-loop of A and C subunits showed notable conformational changes or movements compared to the FG-loop of B on a nanosecond timescale (marked in gray in Fig.~\ref{fig:Capsid}c). The dehydration primarily affects the FG-loop of A and C~\cite{brodmerkel2022stability,perkett2016allosteric}, suggesting a strong role for water molecules in stabilising the extended form of the FG-loop around the 3-fold pore. In addition, mass spectrometry observations hint that a section of the internal RNA stabilises the A/B dimers of the capsid~\cite{knapman2010determining}.

Based on these observations, we formulate a hypothesis that due to the high mobility of the FG-loops of A and C under dehydrating conditions, the FG-loops around the 3-fold pore contract upon losing stabilising waters and the C/C dimer shifts towards the centre of the virus. We utilised the positions of A, B, and C subunits from the asymmetric unit of the 2MS2 crystallographic model as a starting point, then adjusted the position of the C subunit (by translation and rotation) to form a new capsid assembly and minimised the energy of the entire capsid model in vacuum conditions. This procedure was iterated until we obtained a stable capsid shell model which also fit our SPI electron density map, shown in Fig.~\ref{fig:Capsid}d. Figure~\ref{fig:Capsid}e shows the full capsid with the modelled pore structure. The left half shows the low-pass filtered electron density map showing a remarkable similarity to the experimental map in Fig.~\ref{fig:SPI}d. 

\section{Discussion}

The structural response of viruses to a dehydrating environment is an important, and somewhat understudied question, limited by the inability to study these systems \emph{in situ} under these conditions. Single particle imaging using XFELs provides a unique opportunity to probe the structures of these viral capsids while they are dehydrating in an aerosol stream. With the use of machine learning tools to classify the whole ensemble of observed particles, one can observe complex conformational trajectories which would be hidden with other ensemble-averaged measurements. The femtosecond XFEL pulses allow one to temporally freeze the structural transitions and observe non-equilibirum, intermediate structures that occur during dehydration.

In this work, we apply this method to MS2 bacteriophage capsids, where we observe 3D structures ranging from a well-hydrated particle with a liquid envelope down to a dehydrated structure, with a different capsid morphology. Not only do we see these endpoints, but also a large number of intermediate conformations which break icosahedral symmetry, providing clear evidence for a site-specific transformation rather than a capsid-wide concerted change.

While this study is limited to a moderate resolution, the large scale changes in this system are already clearly apparent. Upcoming technical improvements promise to push this resolution barrier to sub-nm levels~\cite{yenupuri2024helium,ayyer2020reference}. This work also opens up the possibility of studying this important question for aerosol-transmitted pathogenic viruses.

\section{\label{sec:methods}Methods}

\subsection{Sample Preparation.}
E. coli strain C-3000 (\textit{ATCC 15597}) was cultured in volumes of 50 ml at 37°C with shaking at 150 rpm. Shaking was reduced to 90 rpm when the exponential growth phase was reached, and the culture was infected with \SI{100}{\micro\liter} MS2 (\SI{2.9}{\milli\gram\per\milli\liter}, $\varepsilon_{280}=$ \SI{3.86}{\milli\gram\per\milli\liter}) (\textit{ATCC 15597-B1}) and \SI{100}{\micro\liter} CaCl$_2$ (1 M). Incubation was stopped when the cells were lysed (c. 3 hours). One milliliter of the lysate and \SI{800}{\micro\liter} CaCl$_2$ was used to infect \SI{400}{\milli\liter} of exponential phase growth culture of E. coli. Incubation was carried out with shaking at 90 rpm until the cells were lysed (ca. 5 hours). The lysate was precipitated using 10\% (w/v) PEG 6000 and 0.5 M NaCl over 48 hours at \SI{4}{\celsius}. 

After precipitation, the suspension was centrifuged at \SI{10000}g for 30 min. The pellet was re-suspended in \SI{30}{\milli\liter} 0.01 M Tris, pH 7.5 (containing 0.1 M NaCl, 0.1 mM MgCl$_2$, and 0.01 mM EDTA). Stirring was carried out for 1 hour at room temperature until complete re-suspension. Next, the suspension was incubated at \SI{37}{\celsius} with shaking at 120 rpm after adding 1.5 mg lysozyme, \SI{300}{\micro\liter} MgCl$_2$ (1 M), and \SI{10}{\micro\liter} Benzonase. After incubation, the suspension was centrifuged at \SI{8000}g for 30 min. The supernatant was precipitated using 10\% (w/v) PEG 6000 and 0.5 M NaCl and incubated at \SI{4}{\celsius} overnight. The suspension was centrifuged at \SI{27000}g for 30 min, and the pellet was re-suspended in Tris buffer. The re-suspension was applied to a sucrose gradient (15-50\%) and centrifuged at \SI{40000}g for 18 h at \SI{4}{\celsius}. The sucrose in the collected band fractions was removed by repetitive concentration and dilution steps with Tris buffer using an Amicon Ultra Centrifugal Filter (100 kDa cutoff).

Prior to cryo-EM grid preparation and sample injection at the XFEL, the Tris buffer of the sample was exchanged to a buffer containing 0.2 mM sodium citrate and 5 mM ammonium acetate using a PD Minitrap G-25 column (Cytiva). The sample concentration was adjusted to $\sim2\times10^{15}$ particles/mL (or $\sim$12 mg/mL) for both experiments.

\subsection{Cryo-EM Structure Determination.}
An aliquot (3 $\mu$L) of MS2 virions was deposited onto freshly glow-discharged, 300 mesh R2/2 Quantifoil grids, followed by 3 s of blotting at 4\textdegree C and 95\% humidity using a Vitrobot Mark IV instrument (ThermoFisher Scientific). The blotted grid was plunge-frozen into a 37:63 (v/v) liquid ethane/propane mixture. Images were acquired using a Talos Arctica microscope (ThermoFisher Scientific) operated at 200 kV and equipped with a Falcon 3EC detector (ThermoFisher Scientific). A total of 861 movies were recorded using the EPU software (ThermoFisher Scientific) in integration mode at a nominal magnification of $\times$92,000, yielding a final pixel size of 1.58 \AA\textsuperscript{2}. Each movie had a total dose of 36e\textsuperscript{-}/\AA\textsuperscript{2} over 39 frames.

Image processing was performed using cryoSPARC~\cite{punjani2017cryosparc}. Drift and beam-induced motions were corrected using patch motion correction, and the contrast transfer function (CTF) was estimated using patch CTF estimation. The micrographs were inspected and curated using the manually curated exposures job, from which 622 micrographs were accepted for further processing. Blob picking was used to pick 60,861 particles, of which 47,546 remained after two rounds of 2D classification. Two classes out of four from \emph{ab initio} reconstruction and heterogeneous refinement (C1 symmetry) had apparent density for both the capsid and the A protein. The particles from these two classes (22,592) were selected for homogenous refinement (C1), where a 4.9 \AA\ resolution map was obtained as estimated by the Fourier shell correlation (FSC) = 0.143 criterion (see Supplementary Fig.S1). 

\subsection{SPI Data Collection.}
Data was collected at the SPB/SFX (single particles, clusters and biomolecules \& serial femtosecond crystallography) instrument~\cite{mancuso2019single} of the European XFEL using \SI{6}{\kilo\electronvolt} photons focused into a $250\times$\SI{250}{\nano\meter\squared} spot. Individual x-ray pulses were generated with \SI{3.8}{\milli\joule} of energy on average ($3.94 \times 10^{12}$ photons/pulse). The pulses were delivered in 352-pulse trains with an intra-train repetition rate of 1.1 MHz and trains arriving every 0.1 s, leading to a maximum data collection rate of 3520 frames/second. A detector built specifically for this burst mode operation, the Adaptive Gain Integrating Pixel Detector (AGIPD)~\cite{henrich2011adaptive}, was placed 700~mm downstream of the interaction region to collect the diffraction patterns for each pulse individually up to a scattering angle of \SI{13}{\degree} at the corner of the detector.

\subsection{Details of $\beta$-VAE Network.}
The $\beta$-VAE consisted of an \textit{encoder} and a \textit{decoder} neural network, to encode information into a lower dimension and retrieve it back respectively. The encoder encodes diffraction data (in this case, 2D intensity models) generating a low-dimensional latent vector, $Z$, for each input pattern $\mathbf{X}$. The encoder parameterises this distribution with a mean $\mu(\mathbf{X})$ and a variance $\sigma(\mathbf{X})$. During training, this distribution is sampled from a normal distribution $\mathcal{N}(\mu(\mathbf{X}), \sigma(\mathbf{X}))$ before being passed to the decoder, which introduces stochasticity, improving robustness and ensuring smoothness of the latent space. 

The network is trained and optimised by minimising a loss function, combining mean square error as a reconstruction loss and Kullback-Leibler (KL) divergence loss as a regularisation term, which discourages a too-sharp latent space. In our case, the optimized $\beta$-VAE had $\beta = 0.5$, with the latent space dimension of $Z = 2$ (see Supplementary Section II for details).

\textit{Pre-processing:} The initial 2D intensities from Dragonfly have dimensions of 503 $\times$ 503 pixels. Preprocessing steps were applied to enhance relevant features and reduce computational redundancy. Given the highly sampled nature of the data and minimal scattering signal at high \(q\), the size was reduced to 171 $\times$ 171 through downsampling and cropping. Additionally, background normalisation was performed by subtracting the mean at high \(q\) and dividing by the mean at low \(q\). Considering that diffraction patterns of compactly supported objects are primarily dominated by low \(q\) signal, to appropriately weight higher \(q\) shape information, the 2D intensities were divided by the radial average intensity over the whole dataset before inputting them into the network. This weighting was then reverted when generating the 3D Fourier volumes. This approach optimises computational efficiency by focusing solely on relevant information in the diffraction data, where distinctive features are evident.

\textit{Network Parameters:} The encoder network comprises a series of convolutional layers, specifically three Conv2d layers that increase in channel depth from 8 to 32, followed by a sequence of linear layers reducing the dimensionality to a latent space dimension Z. Conversely, the decoder utilizes a symmetrical setup starting from the latent dimension Z, expanding through linear layers, and then upscaling spatial dimensions through three ConvTranspose3d layers, ultimately reconstructing the input data. 
Other optimized hyperparameters of the $\beta$-VAE include a batch size of 32 and a learning rate of $10^{-4}$ and wight decay of $10^{-5}$ for Adam optimizer~\cite{kingma2014adam} . Refer to Supplementary Table I for network architecture parameter values.

\subsection{Analysis workflow details}
In order to learn the relevant structural latent variables, we include the quaternion representation of the estimated orientation ($\Omega$) for every 2D intensity model in the dense layers of the encoder network. In the decoding process, after sampling the latent vector, the decoder reconstructs a 3D Fourier volume. This reconstructed volume is then sliced at $\Omega$ to retrieve the input intensities as reconstructed output data. 

As an initial estimate, the orientation of each of these 2D intensity class averages was determined against a single 3D Fourier volume of the icosahedrally symmetrized MS2 bacteriophage from the conventional SPI reconstruction (Fig.~\ref{fig:SPI}d). These orientation estimates were incrementally updated using a so-called Local Optimizer, which works as follows. After a given epoch, each input data frame was used to generate a 3D Fourier volume using a single pass through the VAE. This volume was sliced multiple times, using orientations which were slightly different from the current estimate (standard deviation of \SI{5}{\milli\radian} or \SI{0.3}{\degree}). The updated orientation for this frame was chosen to be the one which maximised the Pearson correlation coefficient with the data (see Supplementary Section III for details). This pipeline is shown schematically in Fig.~\ref{fig:VAE}b.

For this dataset, the $\beta$-VAE was trained over a total of 2000 epochs. In the first 1000 epochs, the Local Optimizer was turned off, and icosahedral symmetric orientation estimates were fed
, allowing the VAE to learn features from the dataset and stabilise itself. In the later 1000 epochs, the orientations were updated. 

\subsection{Phase Retrieval.}
The electron densities were reconstructed through a 3D iterative phase retrieval method applied to the full-resolution intensity volume of the MS2 bacteriophage. The procedure was almost identical to the pipeline discussed in \onlinecite{ayyer2019low}. Fig.~\ref{fig:SPI}d illustrates the reconstructed electron density obtained for a dehydrated phage. In Supplementary Fig.~S3, the phase retrieval transfer function (PRTF) metric, evaluating the reproducibility of retrieved phases based on 128 independent phasing runs for both icosahedral and octahedral structure of MS2 capsid.

The electron density reconstruction from the background-subtracted intensity distribution involved a hybrid approach employing the error reduction (ER) algorithm and the difference map (DM) algorithm. Each phasing run consisted of 400 iterations, comprising 100 ER iterations followed by 200 DM iterations, and concluding with 100 additional ER iterations. The support was updated after each iteration using  a smoothing and thresholding procedure, with the strongest \SI{40000} voxels retained in the support.

The phase retrieval process for the reconstructed Fourier volumes by the decoder network for various trajectory points involved 16 random model starts. The number of voxels for the support was determined based on the estimated diameter size and fringe counts in the 3D Fourier volumes. The electron density maps were visualised with radial colouring to depict structural variations.

The density map, crystal structure and SPI densities were visualised using the Chimera software~\cite{pettersen2004ucsf}.

\subsection{Particle Size Determination.}
We utilised spherical particle fitting on the Fourier volumes reconstructed by the decoder network of $\beta$-VAE. This process involved computing the radial average of the volumes and fitting them with the Fourier model of a spherical particle. This analysis yielded an estimation of the diameter of the MS2 phages during shape-phase transition. The Fourier model for a spherical particle is described by the function \(S(q, d)\):
\[ 
S(q) \propto d^6 \left(\frac{\sin(\pi q d) - \pi q d \cdot \cos(\pi q d)}{(\pi q d)^3}\right)^2 
\]
where $d$ denotes the diameter of the particle and $q$ is defined with the crystallography convention. The size distribution of MS2 bacteriophage is shown in Supplementary Fig.S5.

\subsection{MD Simulation}
We employed the Gromacs package~\cite{lindahl_2022_5938884} for our simulations, utilising the OPLS-AA force field~\cite{brodmerkel2022stability} to investigate the A/B and C/C dimers in vacuum conditions. The initial configurations were based on the 2MS2 PDB structure~\cite{golmohammadi1993refined}. To achieve a total charge of +10e for the dimers, we protonated specific aspartic and glutamic acid residues within each subunit~\cite{knapman2010determining}, adhering to a well-established protocol~\cite{brodmerkel2022stability}. Subsequently, the structures underwent a steepest descent energy minimisation followed by a brief equilibration at 300K, without the application of periodic boundary conditions or pressure coupling, to simulate vacuum conditions. Protein dynamics were monitored over a \SI{20}{\nano\second} period, with all parameters maintained in alignment with the established protocol~\cite{brodmerkel2022stability}.

\bibliography{references}

\noindent \textbf{Acknowledgments}
We acknowledge European XFEL in Schenefeld, Germany, for provision of X-ray free-electron laser beamtime, proposal 2734, at SPB/SFX SASE1 and would like to thank the staff for their assistance. This work is supported by the Cluster of Excellence-CUI: Advanced Imaging of Matter' of the Deutsche Forschungsgemeinschaft (DFG) - EXC 2056 - project ID 390715994, the Swedish Research Council (2018-00234 and 2019-06092) and the Carl Tryggers Stiftelse f{\"o}r Vetenskaplig Forskning (CTS 19-227). We thank Ilme Schlichting for useful discussions.\\

\noindent \textbf{Authors Contributions}
K.A. conceived the project. Samples were prepared by A.Mu.. Cryo-EM data collection and analysis was performed by A.Mu. with the assistance of C.S.. All authors except C.S. performed the XFEL experiment. P.M. performed the molecular dynamics simulations. The diffraction data was analysed by A.Ma., S.Z. and K.A. with the help of P.M. and T.W.. K.A., A.Ma. and P.M. wrote the manuscript with input from all authors.\\

\noindent \textbf{Competing financial interests}
The authors declare no competing financial interests.\\

\noindent \textbf{Code availability}
The code for the $\beta$-VAE is available at \url{https://github.com/AyyerLab/StrucNN/tree/master/ms2vae}. The \emph{Dragonfly} source code is available at \url{https://github.com/duaneloh/Dragonfly}.

\end{document}


\setcitestyle{super}

\title{Supplementary Information: \\Observation of Aerosolization-induced Morphological Changes in Viral Capsids}

\author{Abhishek~Mall}
\affiliation{Max Planck Institute for the Structure and Dynamics of Matter, 22761 Hamburg, Germany}

\author{Anna~Munke}
\affiliation{Center for Free Electron Laser Science, Deutsches Elektronen Synchrotron (DESY), 22607 Hamburg, Germany}
\affiliation{Laboratory of Molecular Biophysics, Department of Cell and Molecular Biology, Uppsala University, Uppsala, SE-75124, Sweden}

\author{Zhou~Shen}
\affiliation{Max Planck Institute for the Structure and Dynamics of Matter, 22761 Hamburg, Germany}

\author{Parichita~Mazumder}
\affiliation{Max Planck Institute for the Structure and Dynamics of Matter, 22761 Hamburg, Germany}

\author{Johan~Bielecki}
\affiliation{European XFEL, Holzkoppel 4, 22869 Schenefeld, Germany}

\author{Juncheng~E}
\affiliation{European XFEL, Holzkoppel 4, 22869 Schenefeld, Germany}

\author{Armando~Estillore}
\affiliation{Center for Free Electron Laser Science, Deutsches Elektronen Synchrotron (DESY), 22607 Hamburg, Germany}

\author{Chan~Kim}
\affiliation{European XFEL, Holzkoppel 4, 22869 Schenefeld, Germany}

\author{Romain~Letrun}
\affiliation{European XFEL, Holzkoppel 4, 22869 Schenefeld, Germany}

\author{Jannik~L{\"u}bke}
\affiliation{Center for Free Electron Laser Science, Deutsches Elektronen Synchrotron (DESY), 22607 Hamburg, Germany}

\author{Safi~Rafie-Zinedine}
\affiliation{European XFEL, Holzkoppel 4, 22869 Schenefeld, Germany}
\affiliation{Institute of Biomaterials and Biomolecular Systems, University of Stuttgart, Pfaffenwaldring 57, 70569 Stuttgart, Germany}

\author{Adam~Round}
\affiliation{European XFEL, Holzkoppel 4, 22869 Schenefeld, Germany}

\author{Ekaterina~Round}
\affiliation{European XFEL, Holzkoppel 4, 22869 Schenefeld, Germany}

\author{Michael~R{\"u}tten}
\affiliation{Universit{\"a}t Hamburg, Institute of Physical Chemistry, Grindelallee 117 20146 Hamburg, Germany}

\author{Amit~K.~Samanta}
\affiliation{Center for Free Electron Laser Science, Deutsches Elektronen Synchrotron (DESY), 22607 Hamburg, Germany}
\affiliation{The Hamburg Center for Ultrafast Imaging, Universität Hamburg, 22761 Hamburg, Germany}

\author{Abhisakh~Sarma}
\affiliation{European XFEL, Holzkoppel 4, 22869 Schenefeld, Germany}

\author{Tokushi~Sato}
\affiliation{European XFEL, Holzkoppel 4, 22869 Schenefeld, Germany}

\author{Florian~Schulz}
\affiliation{Universit{\"a}t Hamburg, Institute of Physical Chemistry, Grindelallee 117 20146 Hamburg, Germany}

\author{Carolin~Seuring}
\affiliation{Centre for Structural Systems Biology (CSSB), Notkestraße 85, 22607 Hamburg, Germany}

\author{Tamme~Wollweber}
\affiliation{Max Planck Institute for the Structure and Dynamics of Matter, 22761 Hamburg, Germany}

\author{Lena~Worbs}
\affiliation{Center for Free Electron Laser Science, Deutsches Elektronen Synchrotron (DESY), 22607 Hamburg, Germany}
\affiliation{Department of Physics, Universität Hamburg, Luruper Chaussee 149, 22761 Hamburg, Germany}

\author{Patrik~Vagovic}
\affiliation{European XFEL, Holzkoppel 4, 22869 Schenefeld, Germany}

\author{Richard~Bean}
\affiliation{European XFEL, Holzkoppel 4, 22869 Schenefeld, Germany}

\author{Adrian~P.~Mancuso}
\thanks{Present address: Diamond Light Source, Harwell Science and Innovation Campus, Didcot, Oxfordshire, OX11 0DE, UK}
\affiliation{European XFEL, Holzkoppel 4, 22869 Schenefeld, Germany}
\affiliation{Department of Chemistry and Physics, La Trobe Institute for Molecular Science, La Trobe University, Melbourne, VIC, 3086, Australia}

\author{Ne-Te~Duane~Loh}
\affiliation{National University of Singapore (NUS), Dep. of Physics / Fac. of Science , 2 Science Drive 3 Singapore 117542 , Singapore}

\author{Tobias~Beck}
\affiliation{Universit{\"a}t Hamburg, Institute of Physical Chemistry, Grindelallee 117 20146 Hamburg, Germany}

\author{Jochen~K{\"u}pper}
\affiliation{Center for Free Electron Laser Science, Deutsches Elektronen Synchrotron (DESY), 22607 Hamburg, Germany}
\affiliation{The Hamburg Center for Ultrafast Imaging, Universität Hamburg, 22761 Hamburg, Germany}
\affiliation{Department of Physics, Universität Hamburg, Luruper Chaussee 149, 22761 Hamburg, Germany}
\affiliation{Department of Chemistry, Universität Hamburg, 20146 Hamburg, Germany}

\author{Filipe~R.N.C.~Maia}
\affiliation{Laboratory of Molecular Biophysics, Department of Cell and Molecular Biology, Uppsala University, Uppsala, SE-75124, Sweden}
\affiliation{NERSC, Lawrence Berkeley National Laboratory, Berkeley, CA, 94720, United States}

\author{Henry~N.~Chapman}
\affiliation{Center for Free Electron Laser Science, Deutsches Elektronen Synchrotron (DESY), 22607 Hamburg, Germany}
\affiliation{Department of Physics, Universität Hamburg, Luruper Chaussee 149, 22761 Hamburg, Germany}
\affiliation{The Hamburg Center for Ultrafast Imaging, Universität Hamburg, 22761 Hamburg, Germany}

\author{Kartik~Ayyer}
\email{kartik.ayyer@mpsd.mpg.de}
\affiliation{Max Planck Institute for the Structure and Dynamics of Matter, 22761 Hamburg, Germany}
\affiliation{The Hamburg Center for Ultrafast Imaging, Universität Hamburg, 22761 Hamburg, Germany}

\maketitle


\clearpage
\section{Cryo-EM FSC curves}

\begin{figure}[h!]
\begin{tabular}{cc}
\includegraphics[scale=0.75]{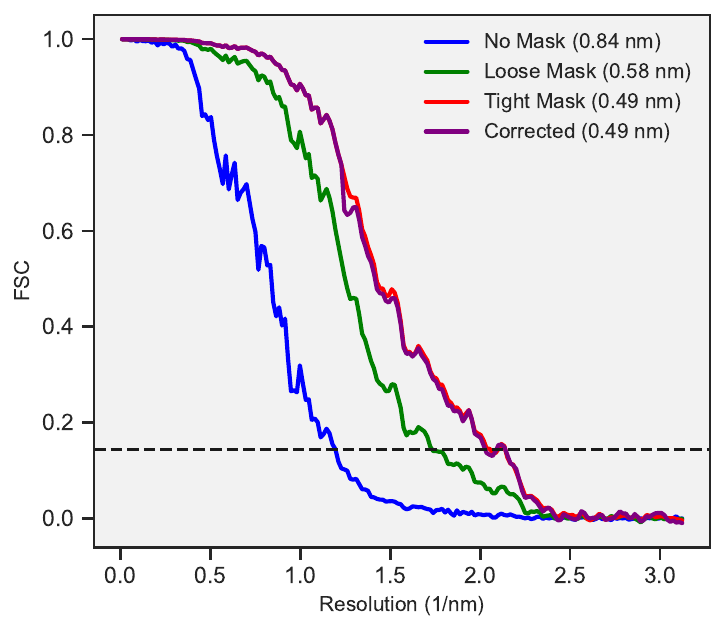} \\
\end{tabular}
\caption{FSC curve of the cryo-EM reconstruction generated using cryoSPARC~\cite{punjani2017cryosparc} for the cryo-EM reconstruction of the MS2 capsid from the same batch as used for the X-ray SPI experiment. The average resolution of 0.49 nm was estimated based on FSC = 0.143 threshold~\cite{rosenthal2003optimal} (black dashed line).}
\label{fig:FSC} 
\end{figure}

\clearpage
\section{Classification, Discrete Heterogeneity and Polymorphism}

\begin{figure*}
\begin{tabular}{c}
\includegraphics[scale=1]{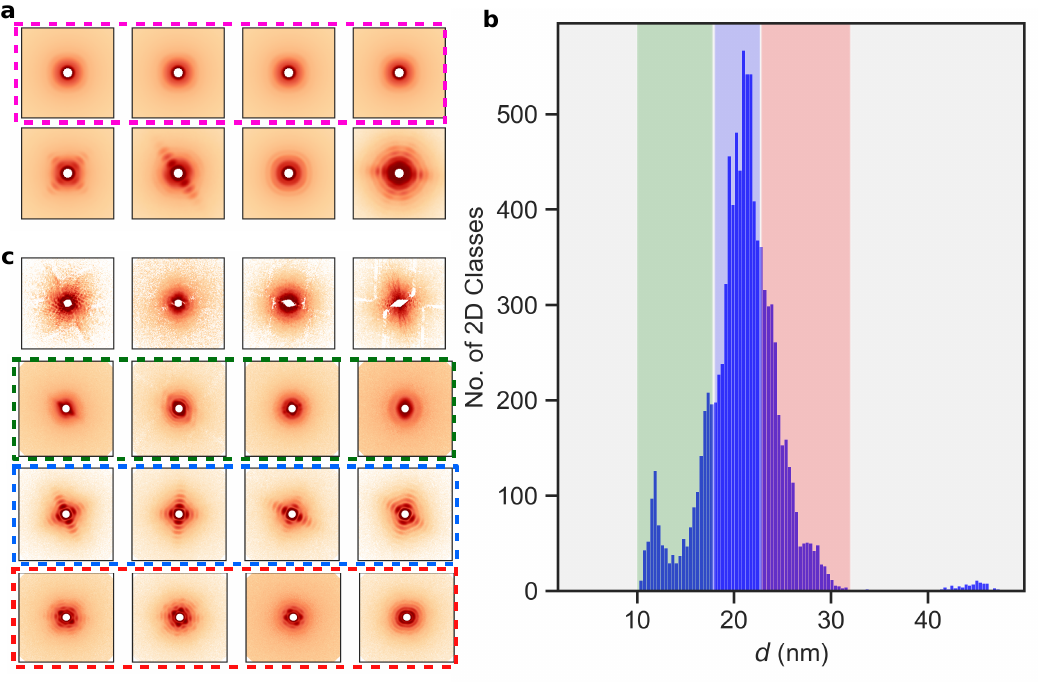} \\
\\[6pt]
\end{tabular}
\begin{tabular}{c}
\includegraphics[scale=1]{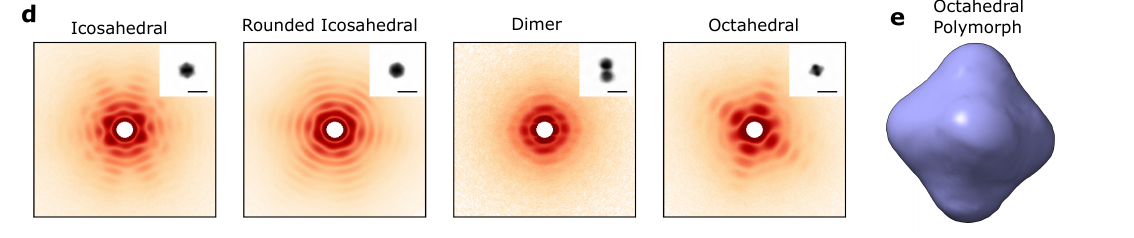} \\
\\[6pt]
\end{tabular}
\caption{\textbf{Classification, Discrete Heterogeneity, and Polymorphism}. (a) Examples from the 50 2D intensity models obtained from the initial EMC classification. The 2D intensity models in the dashed pink grid were rejected prior to generating the training dataset for the $\beta$-VAE. (Bottom row) Examples of the 2D intensity models used for the dataset generation. (b) A histogram of the fitted diameter (nm) values for the 2D intensity classes dataset (10,000 classes). Different structure types of the samples are manually marked in the distribution as icosahedral (red), octahedral (blue), and others, including dimers, outliers, etc. (green). (c) The corresponding examples of 2D intensity models for each structure class are shown in the grid. (Top row) Examples of classes with panel gaps and detector artifacts, which had fitted diameters $>40$ nm. (d) Examples from the diffraction dataset of 2D intensity models and corresponding electron density projections (inset) via phase retrieval. The structural diversity in the diffraction dataset includes icosahedral, rounded icosahedral, dimers and octahedral particles; the scale bar is 30 nm. (e) 3D structure of the MS2 capsid reconstructed from the octahedral data (resolution 6.1 nm). The capsid structure exhibits an octahedral shape — a known form of polymorphism in MS2 phages.}
\label{fig:Objects} 
\end{figure*}

\textit{Classification} : The first step of data classification was the generation of average two-dimensional (2D) classes in the detector plane from the full diffraction dataset using the 2D classification procedure implemented in Dragonfly~\cite{ayyer2016dragonfly,ayyer20213d}. This process employs a modified EMC algorithm to classify all frames into a specified number of averages (models, termed classes in Dragonfly). We began by classifying the dataset into 50 2D classes. Examples of these classes are shown in Fig.~\ref{fig:Objects}a. The 2D classes corresponding to very weak hits (pink dashed grid) were excluded at this stage. 

To obtain the training dataset, a bootstrapping method was employed by running the 2D EMC reconstruction 100 times, each with 100 models, using a random subset of 20$\%$ of the frames (from \num{170355} diffraction frames) each time, resulting in \num{10000} 2D intensity models. Size filtering was then applied to the dataset by fitting a spherical object Fourier model to the radial average of the intensity, resulting in a size distribution of 2D intensity models (Fig.~\ref{fig:Objects}b). By comparing the 2D models and their locations in the distribution, we qualitatively divided the space into three groups, as shown in Fig.~\ref{fig:Objects}b. In the figure, red denotes icosahedral, blue denotes octahedral, and green denotes contaminants, including outliers and dimers. The corresponding example samples of 2D intensities for different groups are shown in Fig.~\ref{fig:Objects}c. The top row shows the classes with panel gaps and detector artefacts, which have fitted diameters greater than 40 nm in the distribution. Among all dataset models, \num{2558} were icosahedral corresponding to \num{79771} diffraction frames. These icosahedral 2D intensity models were used for training the $\beta$-VAE.\\

\textit{Discrete Heterogeneity} : The 2D classification also yielded some interesting structures which had a different symmetry than the icosahedral objects. Figure~\ref{fig:Objects}d shows some of the 2D intensity averages with reasonable intensity contrast. Note that since intensities are always non-negative, the averages from diverse aggregates and contaminants typically generate low-contrast models. The insets show the projected electron densities resulting from 2D phase retrieval. Only patterns belonging to classes like the ``Icosahedral'' class were selected for the reconstruction in Fig.~1d. \\

\textit{Polymorphism} : Along with the rounded icosahedra and dimers, we also obtain patterns with clear octahedral structure. The 3D structure of the octahedral particle was reconstructed without any imposed symmetry from \num{11626} patterns. The reconstructed electron density at \SI{6.1}{\nano\meter} resolution is shown in Fig.~\ref{fig:Objects}e and is 1.53 times lower in volume than the icosahedral structure. MS2 capsids have been reported to assemble with octahedral packing and $T = 3$ quasi-symmetry~\cite{plevka2008crystal,plevka2009structure}. The primary distinction between icosahedral and octahedral structures lies in the presence of four-fold contacts rather than five-fold contacts, potentially resulting in curved interfaces. The octahedral packing results from the fusion of two coat-protein subunits~\cite{peabody1996complementation}. These capsids have been reported to disassemble and reassemble into the octahedral structure during crystallisation~\cite{plevka2008crystal}. To the best of our knowledge, this is the first observation of octahedral MS2 capsids without mutation or different buffer conditions.

\clearpage
\section{Phase retrieval transfer function (PRTF)}
\begin{figure}[h]
\begin{tabular}{cc}
\includegraphics[scale=1]{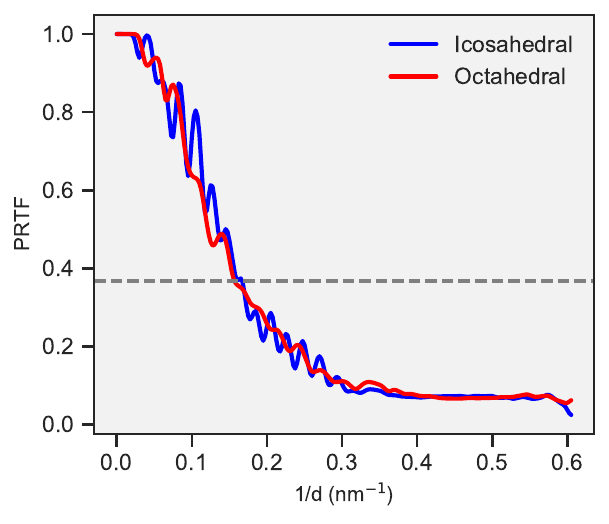} & \includegraphics[scale=1]{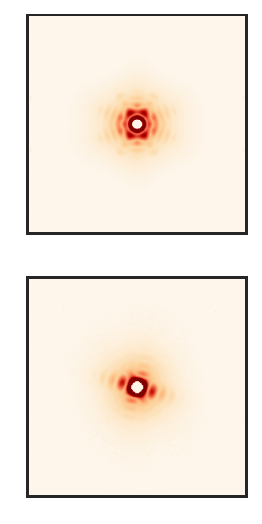} \\
(a) & (b)\\[6pt]

\end{tabular}
\caption{(a) Smoothed phase retrieval transfer function (PRTF) vs \( q \). The solid lines represent the azimuthal average PRTF conventionally used to determine the resolution of the structure. The typical 1/\( e \) cutoff is shown in gray. The resolution at the cutoff for both capsid structures was estimated to be $\approx$ 6.1 nm. (b) Slice at 001-plane through the Fourier volume of the MS2 capsid for icosahedral (top) and octahedral (bottom) structures retrieved using \textit{Dragonfly}~\cite{ayyer2016dragonfly}. Fig.1d $\&$ Fig.~\ref{fig:Objects}e shows the corresponding phased electron density.}
\label{fig:phase} 
\end{figure}

\clearpage
\section{Architecture and Training of $\beta$-VAE}

\begin{figure*}
\begin{tabular}{c}
\includegraphics[scale=1]{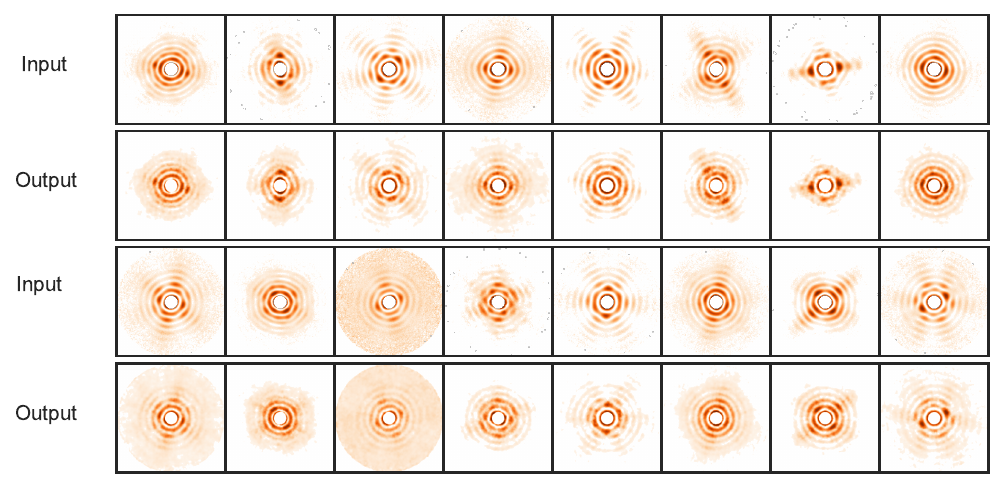} \\
(a) \\[6pt]
\end{tabular}
\begin{tabular}{ccc}
\includegraphics[scale=1]{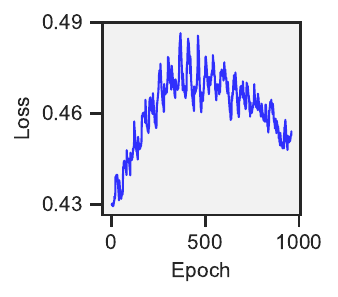} & \includegraphics[scale=1]{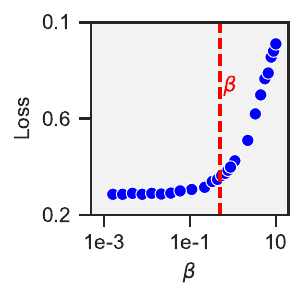} & \includegraphics[scale=1]{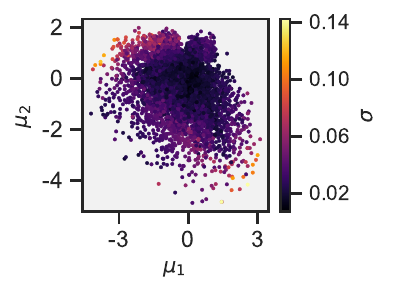}  \\
(b) & (c) & (d)\\[6pt]
\end{tabular}
\begin{tabular}{c}
\includegraphics[scale=1]{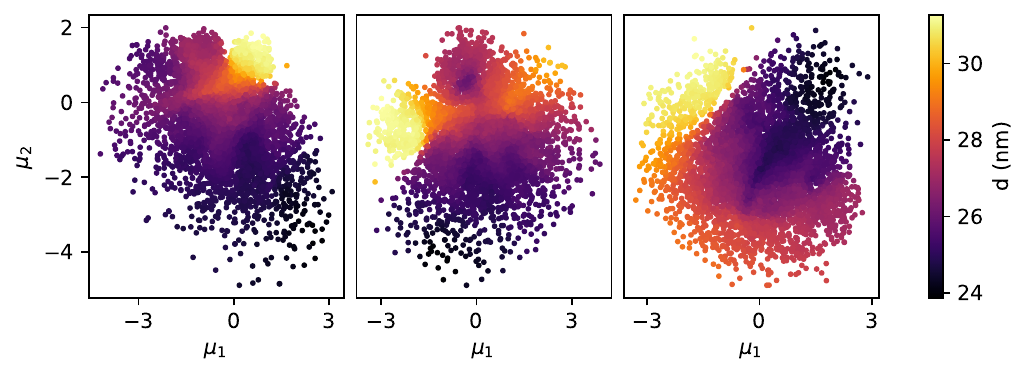} \\
(e) \\[6pt]
\end{tabular}
\caption{(a) Comparison between input 2D intensity data and the corresponding output (reconstruction) by $\beta$-VAE. (b) Loss evolution during $\beta$-VAE training exhibited a decline over the 1000 epochs. However, the decrease was not significant later in training, prompting the decision to terminate further training. The depicted loss encompasses both the Mean Squared Error (MSE) loss for reconstruction and the Kullback-Leibler (KL) divergence loss. (c) $\beta$-VAE Loss \textit{versus} $\beta$ values. The plot illustrates a rise in loss as $\beta$ values increase. The optimal trade-off between minimizing loss and providing sufficient regularization occurred at $\beta$ = 0.5 (red dashed line). (d) Latent space representation of the $\beta$-VAE color labeled with $\sigma = \sqrt{\sigma_{1}^2 + \sigma_{2}^2}$. (e)  Latent space representation of the $\beta$-VAE for different random initializations of weight and bias parameters. The low-dimensional embedding varies due to the random start, converging to nearby minima. However, it maintains the same information for shape and size variations.}
\label{fig:training} 
\end{figure*}

The $\beta$-VAE was trained over a total of 2000 epochs. Fig.~\ref{fig:training} (a) depicts the generative performance of the $\beta$-VAE on 2D intensity data at the final epoch. The majority of prominent features are successfully reconstructed in the output, indicating high-quality reconstruction performance. This highlights the effectiveness of the deep learning model in capturing the fundamental attributes inherent in the input data.

Fig.~\ref{fig:training}(b) illustrates the loss of the $\beta$-VAE over the final 1000 epochs, during which the orientation was updated every 20$^{th}$ epoch before terminating the training. This approach was adopted because the loss stabilized with no significant changes observed. These stable training dynamics suggest efficient convergence of the VAE and optimization of orientation estimates for each 2D intensity model. 

Figure~\ref{fig:training}(c)
illustrates the VAE training to determine the optimal value of $\beta$. The process involved training multiple VAE networks across a range of $\beta$ values from 0 to 10. The optimal value was chosen based on achieving the minimal loss. The selection of $\beta = 0.5$ strikes a balance between smooth disentanglement in the latent space and preservation of reconstruction quality, providing sufficient regularization to prevent overfitting.

Similarly, for \(Z > 2\), there was a reduction in MSE loss; however, this improvement did not reveal any new or distinctive features in the latent space. Conversely, \(Z = 2\) seemed to effectively encapsulate the variations in the dataset. Consequently, a latent dimension of \(Z = 2\) was chosen. 

Figure~\ref{fig:training}(d) shows $\mu_{1}$ and $\mu_{2}$ plot with color-coded label of $\sigma = \sqrt{\sigma_{1}^2 + \sigma_{2}^2}$. The low standard deviation values suggest that the network can effectively extract and learn significant features, which are closely correlated and can be accurately reconstructed with minimal uncertainty.

Figure~\ref{fig:training}(e) shows the latent space representation of the $\beta$-VAE for different random initializations of weights and biases. Specifically, random seeds of 42, 61, and 99 were used, respectively. Although the low-dimensional embeddings appear different due to these random initializations, the latent space consistently captures similar information regarding shape and size variation across all three cases. This demonstrates the robustness of the $\beta$-VAE in retrieving information from diffraction data despite variations in initial conditions.

The architecture parameters of the $\beta$-VAE are mentioned in Table~\ref{tab:vaeparams}.

\begin{table*}[ht]
\centering
\footnotesize 
\begin{tabular}{@{}lllll@{}}
\toprule
\textbf{Network} & \textbf{Layer} & \textbf{Output Size} & \textbf{Weights} & \textbf{Bias} \\ \midrule
Encoder & Conv2d (1, 8) & $H/3 \times W/3 \times 8$ & $5 \times 5 \times 1 \times 8$ & 8 \\
& Conv2d (8, 16) & $H/9 \times W/9 \times 16$ & $5 \times 5 \times 8 \times 16$ & 16 \\
& Conv2d (16, 32) & $H/27 \times W/27 \times 32$ & $5 \times 5 \times 16 \times 32$ & 32 \\
& Linear & 128 & $800 \times 128$ & 128 \\
& Linear & 64 & $128 \times 64$ & 64 \\
& Linear & 8 & $64 \times 8$ & 8 \\
& Linear (mean) & \textit{Z} & $8 \times \textit{Z}$ & \textit{Z} \\
& Linear (log variance) & \textit{Z} & $8 \times \textit{Z}$ & \textit{Z} \\ \midrule
Decoder & Linear & 64 & $\textit{Z} \times 64$ & 64 \\
& Linear & $128 \times 5 \times 5 \times 5$ & $64 \times 128 \times 5 \times 5 \times 5$ & 128 \\
& ConvTranspose3d (128, 64) & $H/3 \times W/3 \times D/3 \times 64$ & $5 \times 5 \times 5 \times 128 \times 64$ & 0 \\
& ConvTranspose3d (64, 32) & $H \times W \times D \times 32$ & $5 \times 5 \times 5 \times 64 \times 32$ & 0 \\
& ConvTranspose3d (32, 1) & $H \times W \times D \times 1$ & $7 \times 7 \times 7 \times 32 \times 1$ & 0 \\ \bottomrule
\end{tabular}
\caption{Architecture of $\beta$-VAE}
\label{tab:vaeparams}
\end{table*}


\begin{figure}
\begin{tabular}{cc}
\includegraphics[scale=0.75]{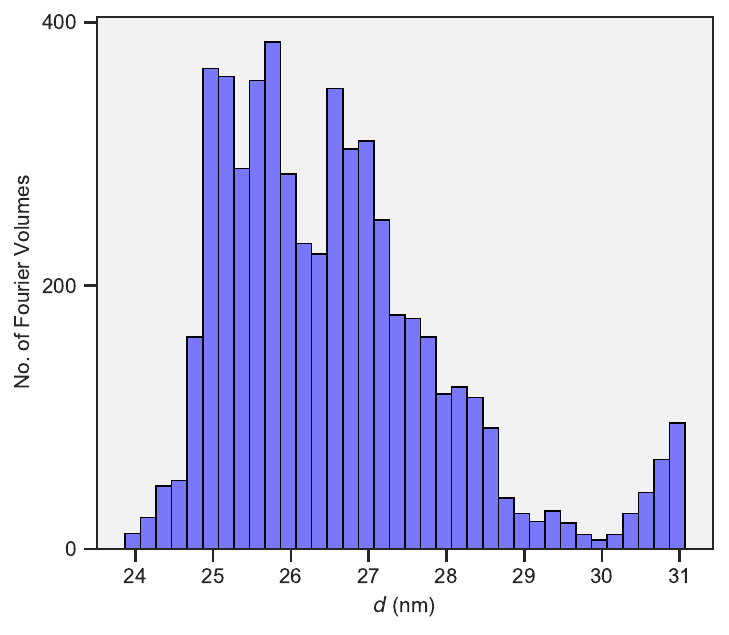} \\
\end{tabular}
\caption{Histogram depicting the particle size distribution, represented by fitted diameter values, \textit{d} (nm). These values were obtained through spherical particle model fitting on the Fourier volumes reconstructed using the \textit{decoder} network of the optimized $\beta$-VAE.}
\label{fig:fit} 
\end{figure}

\clearpage
\section{Local Optimizer for Orientation}
\begin{figure}[h!]
\begin{tabular}{cc}
\includegraphics[scale=0.75]{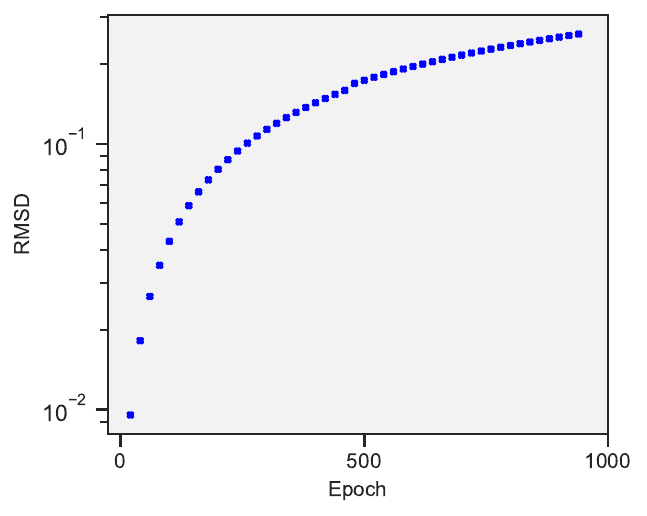} \\
\end{tabular}
\caption{Root mean square deviation (RMSD) \textit{vs} Epoch. The RMSD values were evaluated between the orientation estimates at two consecutive updates of the Local Optimizer. The update is performed every 20$^{th}$ epoch of the training.}
\label{fig:rmsd} 
\end{figure}

The Local optimizer updates the orientation ($\Omega$) every $20^{th}$ epoch during training. To monitor the convergence of orientation estimates, we assess the Root Mean Square Deviation (RMSD) between estimates at consecutive update steps (Fig.~\ref{fig:rmsd}). Convergence is quantified by the measured angle, denoted as $\Theta$, between orientations represented by quaternions at consecutive update steps. $\Theta$ is calculated as:

\[
\Theta = \arccos\left(2 \cdot (\mathbf{q}_1 \cdot \mathbf{q}_2)^2 - 1\right)
\]

where $\mathbf{q}_1$ and $\mathbf{q}_2$ are the normalized quaternions representing orientations.

The RMSD is computed over these angles to measure the average deviation between orientations across update steps of the Local Optimizer. It is determined as:

\[
\text{RMSD} = \sqrt{\frac{1}{n} \sum_{i=1}^{n} \Theta_i^2}
\]

where $n$ is the number of data samples and $\Theta_i$ is the angle between orientations for data sample $i$ at two consecutive epochs.

\clearpage
\bibliography{references}